\documentclass[aps,pra,twocolumn,superscriptaddress,showpacs,10pt]{revtex4-1}
\usepackage{epsfig}
\usepackage{amsfonts,amssymb,amsmath}
\usepackage{amsthm}
\usepackage{color}
\usepackage{mathrsfs}
\usepackage{graphicx}
\usepackage[OT4]{fontenc}
\usepackage{colortbl}
\usepackage[table]{xcolor}

\newcommand{\ket}[1]{\left | #1 \right\rangle}
\newcommand{\bra}[1]{\left \langle #1 \right |}

\renewcommand{\epsilon}{\varepsilon}
\newcommand{\Tr}{\mathrm{Tr}}

\begin{document}

\title{Incompatible local hidden-variable models of quantum correlations}
\date{\today}

\author{Wies{\l}aw Laskowski}
\affiliation{Institute of Theoretical Physics and Astrophysics, University of Gda\'nsk, 80-952 Gda\'nsk, Poland}

\author{Marcin Markiewicz}
\affiliation{Institute of Theoretical Physics and Astrophysics, University of Gda\'nsk, 80-952 Gda\'nsk, Poland}

\author{Tomasz Paterek}
\affiliation{Division of Physics and Applied Physics, School of Physical and Mathematical Sciences, Nanyang Technological University, 637371 Singapore, Singapore}
\affiliation{Centre for Quantum Technologies, National University of Singapore, 3 Science Drive 2, 117543 Singapore, Singapore}

\author{Marcin Wie\'sniak}
\affiliation{Institute of Theoretical Physics and Astrophysics, University of Gda\'nsk, 80-952 Gda\'nsk, Poland}

\begin{abstract}
We show a quantum state with explicit local hidden-variable models for correlations between any fixed number of subsystems
which cannot be extended to a model simultaneously describing correlations between different numbers of subsystems. The explicit models we discuss may involve several settings per observer and the way to disqualify them involves only two settings per party.
\end{abstract}

\pacs{03.65.Ta}

\maketitle

\section{Introduction}
There exist correlations between quantum systems that cannot be explained by any local hidden-variable (LHV) theory.
The simplest scenario that demonstrates this phenomenon involves bipartite entangled quantum states measured with one of two local observables~\cite{BELL,CHSH}.
This original approach of Bell was later extended to correlations between more parties~\cite{GHZ,MERMIN,ZB02,WW} and to correlations between different numbers of subsystems~\cite{BSV12,WBAGV12,WNZ2012,GRAPH1,GRAPH2}.
A natural question arises if there is a correlation Bell inequality that can be violated although all inequalities involving correlations between fixed number of observers are satisfied?

Our main finding is that there exist multiparty states with explicit local hidden-variable models for correlations between any fixed number of subsystems, in a Bell scenario with two settings per party.
Nevertheless these models can be disqualified.
It turns out that they are \emph{incompatible} with each other and cannot be extended to model correlations between various numbers of subsystems.
We present a Bell-like inequality that involves correlations between different numbers of subsystems
which is satisfied by all LHV models and violated by the quantum correlations.

\section{States of interest}
We shall be interested here in states of several two-level quantum systems (qubits).
A general state of $N$ qubits can be represented as follows:
\begin{equation}
\rho = \frac{1}{2^N} \sum_{\mu_1, \dots, \mu_N = 0}^3 T_{\mu_1 \dots \mu_N} \sigma_{\mu_1} \otimes \dots \otimes \sigma_{\mu_N},
\end{equation}
where $\sigma_{\mu_n} \in \{\openone, \sigma_x, \sigma_y,\sigma_z\}$ is the $\mu_n$th local Pauli operator of the $n$th party 
and $T_{\mu_1 \dots \mu_N} \in [-1,1]$ are the components of the extended correlation tensor.
They are given by directly experimentally accessible expectation values $T_{\mu_1 \dots \mu_N} = \Tr[\rho (\sigma_{\mu_1} \otimes \dots \otimes \sigma_{\mu_N})]$.

Consider first so-called Dicke state of $N$ qubits with $e$ excitations
\begin{equation}
\ket{D^e_N} = \frac{1}{\sqrt{{N \choose e}}} \sum_{\pi} \ket{\pi(1 \dots 1 0 \dots 0)},
\label{DEN}
\end{equation}
where $\ket{0}$ ($\ket{1}$) is the eigenstate of $\sigma_z$ local Pauli operator corresponding to $+1$ ($-1$) eigenvalue,
$\pi(1 \dots 1 0 \dots 0)$ denotes a permutation of $e$ ones and $N-e$ zeros, and ${N \choose e}$ gives a number of such permutations.

We focus on the following even mixture of Dicke states for odd number of qubits:
\begin{equation}
\rho_N^e = \frac{1}{2} \ket{D^{e}_N} \bra{D^{e}_N} + \frac{1}{2} | D^{N-e}_N \rangle \langle D^{N-e}_N |.
\label{EVEN_MIX_DICKE}
\end{equation}
This family of states generalizes examples of genuinely multiparty entangled states without $N$-party correlations studied in~\cite{QUANTUM_CORR_WITHOUT_CL,POSTULATES}.
In the Appendix we show that all correlations between odd number of subsystems vanish in the state (\ref{EVEN_MIX_DICKE}), 
whereas all correlations between even number of subsystems are equal to the corresponding correlations of the Dicke states $\ket{D_N^e}$.

This immediately implies that correlations between any odd number of subsystems admit LHV model, namely the model of white noise.
In order to study whether there exists LHV description for correlations between an even number of observers we employ 
the following sufficient condition derived in~\cite{ZB02}.
In a Bell experiment with two settings per observer, correlation functions admit LHV model if
\begin{equation}
\label{wwwzb}
\mathcal{C}_k \equiv \max \sum_{j_1,\ldots, j_k=1}^2 T_{j_1\ldots j_k}^2\leq 1,
\end{equation}
where $k$ is the even number of observers and the maximization is performed over all planes spanned by the Bloch vectors of the local settings.
This optimization can be performed analytically by adapting the method of Ref.~\cite{ZBLW02} and in Table~\ref{TAB_N} we gather the results for several qubits.
In particular, it turns out that for considered states all bipartite correlations admit explicit LHV models of Ref.~\cite{ZB02}, whereas  for a state $\rho_{5}^2$, such explicit models exist for correlations between any number of subsystems\footnote{Actually, the condition \eqref{wwwzb} implies the existence of LHV model for bipartite correlations for any state $\rho_N^e$, since $\mathcal{C}_2 (\rho_N^e)=\frac{8 e^2 (N-e)^2}{(N-1)^2 N^2}\leq\frac{8}{9}$.}. 
Nevertheless, as we show in the next section, these models are incompatible as revealed by a new type of Bell-like inequality.

It is worth mentioning, that due to the results presented in \cite{TDS02}, LHV models for bipartite correlations of states $\rho_N^e$ exist at least for any number of settings of one party and $N-1$ settings of the second party. Indeed, these states are $(1,N-1)$--symmetric extensions of its bipartite reduced states, which according to \cite{TDS02} implies the conclusion.
Furthermore, if $e=(N-1)/2$, then states $\rho_N^e$ can be obtained by tracing out one qubit from a Dicke state $|D_{N+1}^{e+1}\rangle$. Accordingly, for these cases LHV models with $N$ settings for the second party exist. For example in case of $\rho_5^2$ this assures the existence of the model for any number of settings of one party and 5 settings for the second, whereas using the numerical method \cite{STEAM-ROLLER2}
we cannot find a violation of a local realistic model up to ten settings
per side.
Moreover, all models discussed above are more general that the ones coming from condition \eqref{wwwzb}, since they reconstruct all quantum probabilities and not only correlations. 

\renewcommand{\arraystretch}{1.25}

\begin{table}
\begin{tabular}{ c c c c | c c } \hline \hline
State  & ~~~~~~$\mathcal{C}_2$~~~~~~  & ~~~~~~$\mathcal{C}_4$~~~~~~ & ~~~~~~$\mathcal{C}_6$~~~~~~ & ~~$p_{\mathrm{cr}}^{2set}$~~ & ~~$p_{\mathrm{cr}}^{3set}$~~ \\ \hline
$\rho_5^1$ & $\frac{8}{25}=0.32$ & $\frac{33}{25}=1.32$ & $-$ & $0.536$ & $0.477$\\  \hline 
$\mathbf{\rho_5^2}$ & $\mathbf{\frac{18}{25}=0.72}$ & $\mathbf{\frac{24}{25}=0.96}$ & $\mathbf{-}$ & $\mathbf{0.767}$ & $0.746$\\  \hline
$\rho_7^1$ & $\frac{13}{49}\approx 0.27$ & $\frac{25}{49}\approx 0.51$ & $\frac{85}{49}\approx 1.73$ & $0.271$ & -\\ 
$\rho_7^2$ & $\frac{200}{441}\approx 0.45$ & $\frac{32}{147}\approx 0.22$ & $\frac{129}{49}\approx 2.63$ & $0.295$ & -\\ 
$\rho_7^3$ & $\frac{32}{49}\approx 0.65$ & $\frac{864}{1225}\approx 0.71$ & $\frac{256}{245}\approx 1.04$ & $0.508$ & - \\ \hline \hline
\end{tabular}
\caption{\label{TAB_N} Incompatibility of local hidden-variable models.
We present here optimized values of the left-hand side of condition (\ref{wwwzb}) calculated for correlations between two ($\mathcal{C}_2$), four ($\mathcal{C}_4$) and six ($\mathcal{C}_6$)
subsystems of a global system in a state $\rho_N^e$ listed in the first column.
Due to permutational symmetry of the Dicke states the correlations are the same for any particular set of subsystems.
Note that all bipartite correlations admit LHV models of Ref.~\cite{ZB02}.
Moreover, the state $\rho_{5}^2$ (highlighted) admits the model also for four-partite correlations.
The last two columns give critical admixture $p_{\mathrm{cr}}$ such that if $p > p_{\mathrm{cr}}$ there does not exist any LHV model describing quantum probabilities of Bell experiment with two and three settings per observer conducted on a state $p \rho_N^e + (1-p) \rho_{\mathrm{wn}}$, where $\rho_{\mathrm{wn}}$ is a completely mixed state of white noise. 
Therefore, all the discussed mixtures of Dicke states violate some Bell inequality and we present the optimal one for the state $\rho_{5}^2$ in the main text.
In this way we show that although this state admits LHV description on all levels of correlations separately, it does not admit such a description as a whole.}
\end{table}

\section{Bell-type inequality involving correlations between different numbers of subsystems}

We reveal the incompatibility between the LHV model for bipartite correlations and the model for four-partite correlations by showing that these two sets of correlations violate an inequality that combines both of them.

It turns out that it is sufficient to consider only two settings per observer. We now introduce the following Bell inequality:
\begin{eqnarray}
E_{\pi(11110)} + E_{\pi(22220)} + E_{\pi(12220)} && \nonumber \\
 - E_{\pi(21110)} - E_{\pi(11000)} - E_{\pi(22000)} &  \le & 6,
 \label{INEQ}
\end{eqnarray}
where e.g. $E_{\pi(11110)}$ denotes a sum of all correlation functions with indices obtained by permuting elements $(11110)$,
i.e., it is given by the sum of five correlations $E_{\pi(11110)} = E_{11110} + E_{11101} + E_{11011} + E_{10111} + E_{01111}$.
We denote by $E_{kl000}$ correlations between measurement results obtained when the first observer sets his measuring device to the $k$th setting, the second observer sets his apparatus to the $l$th setting, and the remaining observers are not relevant. 
Similarly for correlations between four observers.
Counting all the permutations involved in this inequality, one finds that it is a sum of $70$ terms, $20$ of which are bipartite correlations and $50$ terms being correlations between four subsystems.

The bound of inequality \eqref{INEQ} is easily verifiable on a computer.
The extremal value of the Bell expression on the left-hand side of \eqref{INEQ} is attained for deterministic LHV models,
i.e. the models that perfectly predetermine results of all possible measurements.
In our case of five observers each choosing one of two measurement settings we have altogether ten predetermined results,
$A_1,A_2,\dots, E_1, E_2 = \pm 1$, where $A_1$ is the result the first observer would obtain if he were to measure the first setting,
$A_2$ is the result the first observer would obtain if he were to measure the second setting and so on.
We denote the two possible results of a measurement by $\pm 1$.
Correlation function is defined as expectation value of the product of measurement results
and therefore within a deterministic LHV theory a correlation function is just a product of predetermined results, e.g., $E_{kl000} = A_k B_l$ and $E_{klmn0} = A_k B_l C_m D_n$.
Inserting such products in the Bell expression on the left-hand side of \eqref{INEQ} and checking its value for all $1024$ combinations of predetermined results one finds that the left-hand side attains only three values: $-26, -10$ and $6$. Hence the upper bound of $6$ holds for all LHV correlations.

\section{Quantum violation}
We have chosen the Bell inequality \eqref{INEQ} because in the following sense it is the optimal inequality for the state $\rho_5^2$.
Consider a mixed state $\rho = p \rho_5^2 + (1-p) \rho_{\mathrm{wn}}$, where $\rho_{\mathrm{wn}} = \frac{1}{2^5} \openone$ represents a completely mixed state of no correlations whatsoever and therefore admitting LHV model.
We have verified numerically using software described in Ref.~\cite{STEAM-ROLLER2} that the critical value of $p$ above which the state $\rho$ violates some Bell inequality equals $p_{\mathrm{cr}} = 0.7671$ (see Table~\ref{TAB_N}).
Exactly the same value is found using inequality \eqref{INEQ}.
Accordingly, the highest quantum value of the left-hand side of \eqref{INEQ} is given by $7.8217$.
Almost this maximal violation is observed for a very simple set of measurement settings.
If the same settings described by Bloch vectors
\begin{eqnarray}
\vec s_1 & = & (\cos\tfrac{\pi}{5},- \sin\tfrac{\pi}{5},0), \\ 
\vec s_2 & = & (\cos\tfrac{\pi}{20},\sin\tfrac{\pi}{20},0),
\end{eqnarray}
are chosen by all the observers, the value of the Bell expression (\ref{INEQ}) measured on the state $\rho_5^2$ is given by $7.7831$.

\section{Conclusions and outlook}
We presented a new type of Bell argument that involves correlation functions between different numbers of observers.
This new inequality is shown to be violated by quantum predictions for a class of multiparty entangled states
for which we also show that correlations between any fixed number of subsystems admit LHV models.
We conclude that these models are incompatible and cannot be extended to explain all the correlations of the quantum states.
We hope this research will stimulate experimental demonstration of the incompatibility.

Dicke states with various fidelities have been realized up to six qubits encoded in polarization of photons~\cite{FOUR_PHOTON_DICKE,ENT_SIX_PHOTON_DICKE,SIX_PHOTON_DICKE}.
Note that the state $\rho_5^2$ can be obtained by tracing out one qubit from a six-qubit pure Dicke state with three excitations,
and this method has been used to observe some properties of the state $\rho_5^2$~\cite{ENT_SIX_PHOTON_DICKE}.
However, the measurements performed up to date are not of the sort required by our inequality.
A new experiment is necessary in order to demonstrate the incompatibility.

\section{Acknowledgments}
This work is part of the Foundation for Polish Science TEAM project cofinanced by the EU European Regional Development Fund,
and is supported by the National Research Foundation and Ministry of Education in Singapore.

The contribution of MM is supported within the International PhD Project ``Physics of future quantum-based information technologies'' grant MPD/2009-3/4 from the Foundation for Polish Science and by the University of Gda\'nsk grant 538-5400-0981-12.

WL is supported by the National Centre for Research and Development (Chist-Era Project QUASAR).

MW is supported by the National Science Centre Grant no. N202 208538 and by the Foundation for Polish Science under the HOMING Programme cofinanced by the EU European Development Fund.

TP is supported by the EU program Q-ESSENCE (Contract No. 248095).

\section{Appendix}

\appendix

\section{Correlations of states $\rho_N^e$}

Here we show that states (\ref{EVEN_MIX_DICKE}) have vanishing correlations between an odd number of subsystems and correlations between an even number of subsystems are the same as those of the Dicke state $| D_N^e \rangle$.

Consider $k$ observers performing local measurements on the Dicke state.
The correlations they observe are given by the average value:
\begin{equation}
T_{j_1 \dots j_k 0 \dots 0}(D_N^e) = \langle D_N^e | \sigma_{j_1} \otimes \dots \otimes \sigma_{j_k} \otimes \openone \otimes \dots \otimes \openone | D_N^e \rangle,
\label{DICKE_CORR}
\end{equation}
where $j_n = x,y,z$ and due to permutational symmetry of the Dicke state every set of $k$ observers measures the same correlations.
Anti-Dicke states are obtained by flipping all the qubits in the Dicke states:
\begin{equation}
|D_N^{N-e} \rangle = \sigma_x \otimes \dots \otimes \sigma_x | D_N^e \rangle,
\end{equation}
and therefore their correlations are also given by the right-hand side of (\ref{DICKE_CORR}) but with Pauli operators $\sigma_{j_n}$ replaced by $\sigma_x \sigma_{j_n} \sigma_x$.
Note that $\sigma_x \sigma_{y} \sigma_x = - \sigma_y$ and $\sigma_x \sigma_{z} \sigma_x = - \sigma_z$.
Furthermore, the non-zero correlation tensor components have even number of $x$ and $y$ indices.
Indeed, if the number of $y$ indiced is odd, the action of Pauli operators on the Dicke state produces imaginary global phase and since correlations are real they must vanish.
If the number of $x$ indices is odd, the total number of qubits flipped by the application of Pauli operators is also odd and therefore after the flip the number of excitations is different than before and correlations vanish.
In conclusion, the correlations of anti-Dicke states are either the same or opposite to those of Dicke states and this is decided by the parity of the number of $z$ indices which is the same as the parity of the number of measured systems:
\begin{equation}
T_{j_1 \dots j_k 0 \dots 0}(D_N^{N-e}) = (-1)^k T_{j_1 \dots j_k 0 \dots 0}(D_N^e). 
\end{equation}
This property applied to even mixture of Dicke and anti-Dicke states concludes the proof.


\begin{thebibliography}{99}

\bibitem{BELL}
J. S. Bell,
Physics {\bf 1}, 195 (1974)

\bibitem{CHSH}
J. F. Clauser, M. A. Horne, A. Shimony, and R. A. Holt,
Phys. Rev. Lett. {\bf 23}, 880 (1969).

\bibitem{GHZ}
D. M. Greenberger, M. A. Horne, and A. Zeilinger,
in \emph{Bell's theorem, quantum theory, and conceptions of the universe},
ed. by M. Kafatos (Kluwer, Dordrecht).

\bibitem{MERMIN}
N. D. Mermin,
Phys. Rev. Lett. {\bf 65}, 1838 (1990).

\bibitem{ZB02}
M. \.Zukowski and {\v C}. Brukner,
Phys. Rev. Lett. {\bf 88}, 210401 (2002).

\bibitem{WW}
R. F. Werner and M. M. Wolf,
Phys. Rev. A {\bf 64}, 032112 (2001)

\bibitem{BSV12}
N. Brunner, J. Sharam and T. Vertesi,
Phys. Rev. Lett. {\bf 108}, 110501 (2012).

\bibitem{WBAGV12}
L. E. W\"urflinger, J.-D. Bancal, A. Acin, N. Gisin, and T. Vertesi,
arXiv:1203.4968

\bibitem{WNZ2012}
M. Wie\'sniak, M. Nawareg, and M. \.Zukowski,
arXiv:1112.0951

\bibitem{GRAPH1}
O. G\"uhne, G. T\'oth, P. Hyllus, and H. J. Briegel,
Phys. Rev. Lett. {\bf 95}, 120405 (2005).

\bibitem{GRAPH2}
V. Scarani, A. Acin, E. Schenck, and M. Aspelmeyer,
Phys. Rev. A {\bf 71}, 042325 (2005).

\bibitem{QUANTUM_CORR_WITHOUT_CL}
D. Kaszlikowski, A. Sen(De), U. Sen, V. Vedral and A. Winter,
Phys. Rev. Lett. {\bf 101}, 070502 (2008).

\bibitem{POSTULATES}
C. H. Bennett, A. Grudka, M. Horodecki, P. Horodecki, and R. Horodecki,
Phys. Rev. A {\bf 83}, 012312 (2011).

\bibitem{ZBLW02}
M. \.Zukowski, {\v C}. Brukner, W. Laskowski and M. Wie\'sniak,
Phys. Rev. Lett. {\bf 88}, 210402 (2002).

\bibitem{TDS02}
B. M. Terhal, A. C. Doherty and D. Schwab,
Phys. Rev. Lett. {\bf 90}, 157903 (2003) .


\bibitem{STEAM-ROLLER2}
J. Gruca, W. Laskowski, M. \.Zukowski, N. Kiesel, W. Wieczorek, C. Schmid, and H. Weinfurter,
Phys. Rev. A {\bf 82}, 012118 (2010).

\bibitem{FOUR_PHOTON_DICKE}
N. Kiesel, C. Schmid, G. Toth, E. Solano, and H. Weinfurter,
Phys. Rev. Lett. {\bf 98}, 063604 (2007)

\bibitem{SIX_PHOTON_DICKE}
R. Prevedel, G. Cronenberg, M. S. Tame, M. Paternostro, P. Walther, M. S. Kim, and A. Zeilinger,
Phys. Rev. Lett. {\bf 103}, 020503 (2009).

\bibitem{ENT_SIX_PHOTON_DICKE}
W. Wieczorek, R. Krischek, N. Kiesel, P. Michelberger, G. Toth, and H. Weinfurter,
Phys. Rev. Lett. {\bf 103}, 020504 (2009).

\end{thebibliography}
\end{document}